# Review of Nanolayered Post-transition Metal Monochalcogenides: Synthesis, Properties, and Applications

*Mingyu Yu,[1] Maria Hilse,[2] Qihua Zhang,[2] Yongchen Liu,[1] Zhengtianye Wang,[1,2] Stephanie Law[2,a]*

[1]Department of Materials Science and Engineering, University of Delaware, 201 Dupont Hall, 127 The Green, Newark, Delaware 19716 USA

[2]Department of Materials Science and Engineering and 2D Crystal Consortium Material Innovation Platform, Materials Research Institute, The Pennsylvania State University, University Park, Pennsylvania 16802 USA

[a] Electronic mail: sal6149@psu.edu

**Abstract**

Nanolayered post-transition metal monochalcogenides (PTMMCs) stand out as promising advanced two-dimensional (2D) materials. Beyond inheriting the general advantages associated with traditional 2D materials, they exhibit unique properties, including a wide bandgap range covering the ultraviolet to the mid-infrared spectral ranges, thickness-dependent bandgap behaviors, good nonlinear optical performance, high thermoelectric coefficients, and ferroelectricity. Consequently, these materials hold significant potential in diverse applications such as photodetectors, field effect transistors, thermoelectrics, ferroelectrics, photovoltaics, and electrochemical devices, especially in the manufacturing of nanoscale devices. However, there is still a lack of systematic understanding of the PTMMC family. This study provides a broad overview of the crystal structures, bandgap structures, synthesis methods, physical properties, and state-of-the-art applications of PTMMC materials with a motif of X-M-M-X (M=Ga, In, Ge, Sn; X=S, Se, Te). An outlook for the development trends is emphasized at the end, underscoring the critical importance of this work to the future exploration of nanolayered PTMMCs.

# 1. Introduction

Over the past decades, the outstanding properties of graphene[1] have raised interest in the synthesis and study of two-dimensional (2D) layered materials. The vigorous development of layered materials has triggered revolutionary advances in technologies such as quantum information science, energy conversion, and energy storage, as they easily achieve nanoscale dimensions, their number of layers can easily be controlled, and they possess unique optoelectronic properties, excellent mechanical strength and flexibility, and large surface areas.[2,3] Recently, the development of post-transition metal chalcogenides (PTMCs) has substantially broadened the variety of layered materials. PTMCs are composed of post-transition metals (M) and chalcogens (X=S, Se, Te). Post-transition metals typically refer to elements in the periodic table between transition metals and metalloids, with Al, Ga, In, Tl, Ge, Sn, Pb, Sb, and Bi being the most common (some transition metals, such as Zn, Cd, Hg, may also be considered post-transition metals, but are not used here). There are an extensive number of known PTMC compounds with stoichiometries like $M_2^{III}X^{VI}$,[4,5] $M^{III}X^{VI}$,[4,6-8] $M_2^{III}X_3^{VI}$,[7-22] $M^{IV}X^{VI}$,[6,23] $M^{IV}X_2^{VI}$,[6] and $M_2^{V}X_3^{VI}$ (The superscript Roman characters indicate the group to which the element belongs, and the subscript Arabic numerals indicate the stoichiometric number). The MX subgroup (post-transition metal monochalcogenide, i.e., PTMMC) is of particular interest due to their relatively simple crystal structures, and they are the focus of this review. Some MX crystals, such as GeTe and SnTe, are not classified as layered van der Waals materials. They possess an unconventional metavalent bonding (MVB) mechanism.[24-26] MVB solids often exhibit an unusual combination of properties, including large chemical bond polarizability, large optical dielectric constants, strong light absorption, pronounced lattice anharmonicity, low thermal conductivities, and moderate electrical conductivities (higher than ionic/covalent materials but lower than metals).[24-26]

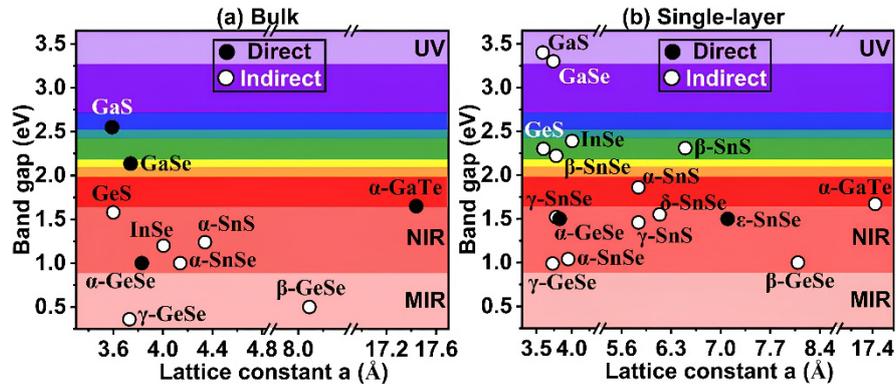

**Figure 1.** Lattice constant vs. band gap plots of the stable layered PTMMC materials in bulk (a) and single-layer (b) forms. Black and white filled circles represent the direct and indirect bandgaps, respectively. The background corresponds to the solar spectrum, where UV, NIR and MIR refer to the ultraviolet, near infrared and mid-infrared regions, respectively.

A major advantage of PTMMCs is their large range of bandgaps, from 0.36 to 3.4 eV,[27-46] as visualized in Figure 1. Moreover, the bandgaps of layered PTMMCs usually have interesting thickness-dependent behaviors: for instance, the direct-to-indirect bandgap transition in GaSe as it moves from bulk to single layer thickness is opposite to the trend observed in transition metal dichalcogenides (TMDs).[9,29,47] Layered PTMMC semiconductors possess additional merits such as high carrier mobility and density as well as p-type electronic properties, providing a wealth of opportunities for developing devices in optics, electronics, and optoelectronics.[31,32,48,49] This review focuses on the state-of-the-art research on layered van der Waals PTMMC materials and emphasizes their properties and applications at the nanoscale. We classify the PTMMCs into four groups: GaX, InX, GeX, and SnX. Each section introduces the material crystal structure, electronic band gap, physical properties, synthesis, and application. Challenges and prospects are discussed at the end of the article.

## 2. Ga Monochalcogenides

### 2.1 Crystal Structure and Band Structure

GaX are layered semiconductors, and their structures can be formulated as $Ga_2^{4+}2X^{2-}$.[50] GaS and GaSe share the same hexagonal crystal structure. Each unit cell is composed of individual quadruple layers X'-Ga-Ga-X' (X'=S, Se) with a height of 7.5 Å for GaS[51,52] and 8.0 Å for GaSe.[29,51] The GaX' single layer has a non-centrosymmetric space group of $D_{3h}$,[29] in which each Ga atom is tetragonally bonded to three chalcogen atoms and one Ga atom, while each chalcogen atom is trigonally coordinated to three Ga atoms. The intralayer bonds are predominately covalent in nature, while the interlayer bonds are van der Waals (vdW) interactions that allow a variety of possibilities for layer stacking, resulting in multiple polytypes. Bulk GaSe crystals have four common polytypes: β, ε, γ, and δ.[6,53] Their crystal configurations are summarized in Table 1 and Figure 2. AB stacking has been found to be more stable than AA' stacking, so ε-GaSe (ABAB) is the most widely studied type, followed by γ-GaSe (ABCABC).[54] Recently, a novel γ'-GaSe with a centrosymmetric space group of $D_{3d}$ and a γ-type stacking sequence has been experimentally observed by Grzonka et al.[29] Its inversion symmetry makes it particularly useful in optoelectronic applications.[55]

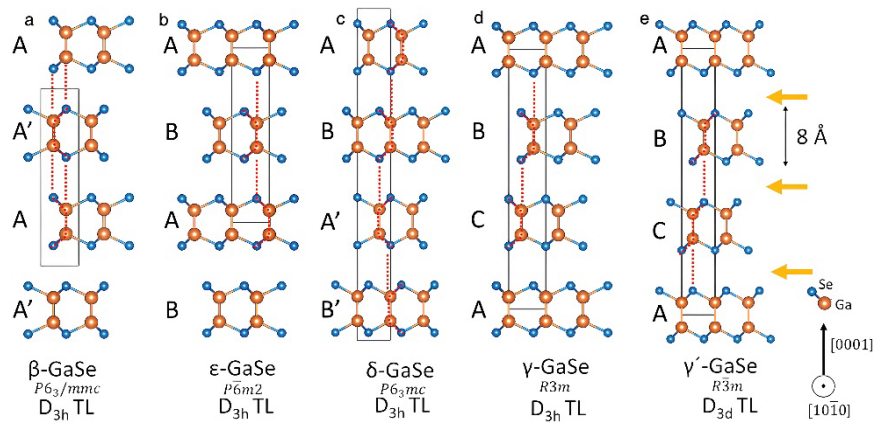

**Figure 2.** Schematic models of (a) β-, (b) ε-, (c) δ-, (d) γ-, and (e) γ'-GaSe from a side view. Yellow arrows indicate vdW gaps. Black rectangles represent unit cells of GaSe. Red dashed lines mark the different geometric patterns and layer alignments. Reprinted with permission

from [29]. Copyright 2021 John Wiley and Sons.

**Table 1.** Summary of the crystallographic properties of known GaSe polytypes.[29,53,54,56-58]

| GaSe Type | Stack order | Space group | | Lattice Parameters | | Interatomic distance (Å) | | |
|---|---|---|---|---|---|---|---|---|
| | | Hermann-Mauguin | Schö-nflies | a=b (Å) | c (Å) | Intra-layer Se-Ga | Intra-layer Ga-Ga | Inter-layer Se-Se |
| β (2H) | AA' | $P6_3/mmc$ | $D^4_{6h}$ | 3.735 | 15.887 | 2.485 | 2.383 | 3.840 |
| ε (2H') | AB | $P\bar{6}m2$ | $D^1_{3h}$ | 3.735 | 15.887 | 2.485 | 2.383 | 3.840 |
| γ (3R) | ABC | $R3m$ | $C^5_{3v}$ | 3.739 | 23.862 | 2.467 | 2.386 | 3.847 |
| δ (4H) | ABA'B' | $P6_3mc$ | $C^4_{6v}$ | 3.755 | 31.990 | 2.463 | 2.457 | 3.880 |

GaS has a unit cell of a=b=3.587 Å and c=15.492 Å.[27] There is only one polytype of GaS, β-GaS, which has a similar $P6_3/mmc$ space group to β-GaSe.[57] The intralayer interatomic distances of S-Ga and Ga-Ga are 2.334 Å and 2.245 Å, respectively, while the interlayer interatomic distance of S-S is 3.768 Å.[27]

GaTe is similarly comprised of 7.47 Å-thick Te-Ga-Ga-Te quadruple layers[50] that can be stacked by vdW forces to form a bulk crystal. There are two phases of GaTe: β-hexagonal[31,59,60] and α-monoclinic.[31,60-62] β-GaTe has a β-GaSe-like structure and is easily transformed to a naturally stable α-GaTe phase.[60,63] α-GaTe has two types of Ga-Ga bonds: two-thirds of them are perpendicular to the layer plane, while the other third are parallel to the layer plane. In contrast, all Ga-Ga bonds are perpendicular to the plane in β-GaTe.[31,48,51] Such a reconstruction in Ga-Ga bonds is caused by the steric occupation of large Te anions,[6] leading to an asymmetric structure as shown in Figure 3. More information about GaTe crystals can be found in Table 2. The conversion from β- to α-GaTe occurs near 500°C.[60]

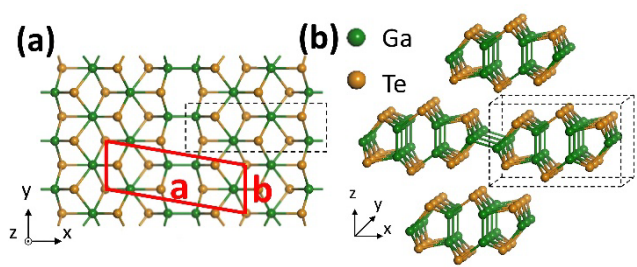

**Figure 3.** Schematic models of bulk α-GaTe (a) from a top view (red frame marks a unit cell, a and b are the lattice parameters shown in Table 2), (b) from a side view. Black dashed boxes

in (a) and (b) indicate the same group of atoms. Reprinted with permission from [64]. Copyright 2016 American Chemical Society.

Table 2. GaTe crystal information.[31,32,50,59,61]

| GaTe Type | Space group | | Lattice Parameters | Number of atoms per unit cell |
|---|---|---|---|---|
| | Hermann-Mauguin | Schönflies | | |
| Hexagonal (β) | $P6_3/mmc$ | $D^4_{6h}$ | a=b=4.06 Å c=16.96 Å | 4 Ga atoms + 4 Te atoms |
| Monoclinic (α) | $C2/m$ | $C_{2h}$ | a=17.44 Å, b=4.07 Å, c=10.46 Å, β=104.5° | 6 Ga atoms + 6 Te atoms |

The bandgap of GaX crystals decreases as the thickness of the crystal increases, saturating when the thickness exceeds tens of layers.[29,31,47] Furthermore, the transition between indirect and direct bandgaps typically happens when the film is ~3-layer thick.[65] Generally, a direct bandgap is ideal for optoelectronic devices, but if the valence band is flat enough (as is the case in GaS, GaSe, and InSe), the films remain optically active despite the indirect bandgap. Figure 4 exhibits simulated band diagrams for GaS and β-GaSe. GaS and β-GaSe both show a "wine bottle" shaped valence band, resulting in a high density of states at the valence band maximum, leading to promising applications in the fields of spintronics and thermoelectrics.[29,66,67] Atomically thin (⩽ 3 layers) GaS and GaSe layers both show indirect transitions, while increasing the number of layers narrows the band gap.[67,68] It is worth noting that in GaSe, the energy difference between the direct and indirect transitions is extremely small, allowing GaSe to behave like a direct bandgap material even when it is extremely thin.[69,70]

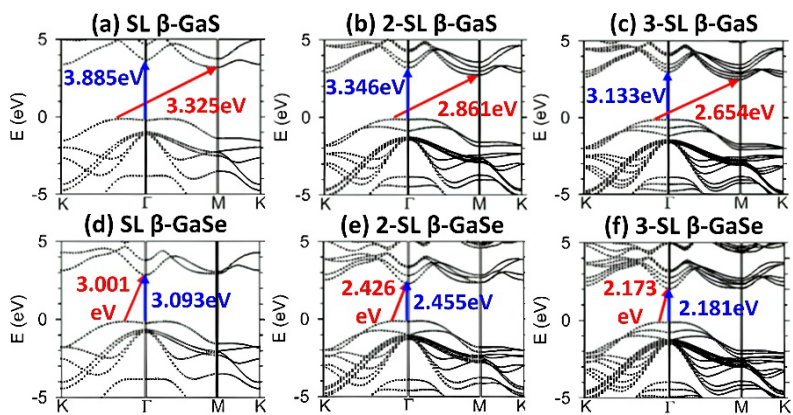

**Figure 4.** Band structures of (a) single-layer (SL), (b) 2-SL, (c) 3-SL β-GaS, (d) SL, (e) 2-SL, and (f) 3-SL β-GaSe. Arrows indicate the indirect (red) and direct (blue) electronic transitions. The values of direct (Γ→Γ) and indirect bandgaps (K-Γ) →M are noted in blue and red, respectively. Reprinted with permission from [28]. Copyright 2015 American Chemical Society.

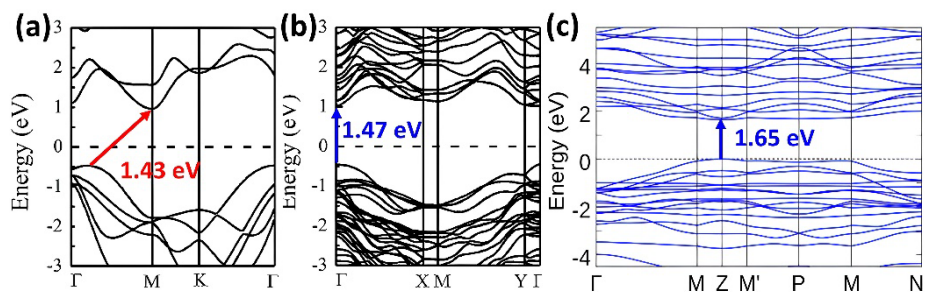

**Figure 5.** Band structures of single-layer (a) β-GaTe and (b) α-GaTe. Reproduced with permission from [71]. Copyright 2019, Springer Nature. (c) Band structure of bulk α-GaTe. Reprinted with permission from [64]. Copyright 2016 American Chemical Society. They were all calculated from a generalized gradient approximation (GGA) with Perdew-Burke-Ernzerhof (PBE) functionals. Arrows indicate indirect (red) and direct (blue) transitions. The results are consistent with other work.[31,51,61]

α-GaTe, has a small direct bandgap and shows strong excitonic absorption and emission even under ambient conditions, making it ideal for high-performance photocatalytic and optoelectronic applications.[71,72] The reduction in symmetry in α-GaTe gives rise to a markedly different electronic structure from that of β-GaTe, as illustrated by Figure 5a,b. Single-layer β-GaTe has an indirect bandgap of 1.43 eV while single-layer α-GaTe has a direct bandgap of 1.47 eV at the Γ point. When α-GaTe is bulk-like, a direct band gap of 1.65 eV is found at the Z point, as shown in Figure 5c. In single-layer α-GaTe, strong dispersion is observed near the top of the valence band along the Γ-X direction, suggesting a low effective mass along this direction. Conversely, along the Y-Γ direction, the dispersion is rather flat, leading to a large carrier effective mass. This combination of flat and dispersive bands near the valence band edge is promising for thermoelectric devices, because a flat band results in a high Seebeck coefficient while a dispersive band improves electrical conductivity.

## 2.2 Electronic and Optical Properties

Carrier type, density, and mobility are among the primary concerns for optoelectronic devices. Multilayer GaS is typically n-type with carrier densities in the range $10^{12}$–$10^{13}$ cm$^{-3}$ and mobilities around 0.1 cm$^2$V$^{-1}$s$^{-1}$, while multilayer ε-GaSe and α-GaTe are both p-type, with carrier densities and mobilities of $10^{14}$–$10^{15}$ cm$^{-3}$ and 0.6 cm$^2$V$^{-1}$s$^{-1}$ for ε-GaSe and $10^{16}$–$10^{18}$ cm$^{-3}$ and 30–40 cm$^2$V$^{-1}$s$^{-1}$ for α-GaTe. The carrier mobilities of their bulk counterparts are larger: 80 cm$^2$V$^{-1}$s$^{-1}$ for GaS and 215 cm$^2$V$^{-1}$s$^{-1}$ for ε-GaSe.[31,73] ε-GaSe and α-GaTe are attractive since it is rare to find p-type conduction in vdW materials. P-type carriers in GaSe and GaTe are attributed to Ga vacancies, while n-type carriers in GaS are due to S vacancies.[59] α-GaTe usually serves as a better candidate for devices than GaS and GaSe due to its higher carrier density, larger mobility, longer lifetime, and unique in-plane anisotropy.[72] The in-plane mobility in α-GaTe varies significantly with orientation: there is a two-order of magnitude increase in resistivity from the $b_{\parallel}$ to $b_{\perp}$ direction.[74]

Layered GaS, GaSe and GaTe materials all exhibit large out-of-plane anisotropy in their mechanical, optical, and electrical properties, a common feature in vdW compounds.[56,59-65,67,69-81] Uniquely, α-GaTe also displays in-plane anisotropy, which has traditionally been attributed to its asymmetric crystal structure. However, recently Lai *et al.*[82] claimed that strong bulk-surface interactions are the true underlying cause. Hexagonal GaSe crystals have garnered particular attention due to their nonlinear optical properties in the infrared (IR) range[83] and their exceptional transparency between 650–18000 nm.[56,77]

**2.3 Synthesis**

In general, synthesis methods can be divided into two categories: top-down and bottom-up approaches. The former starts with bulk materials and then separates thin layers by mechanical exfoliation, chemical intercalation, solution-based sonication, or selective etching. This approach works for most vdW materials, but it has challenges in scalability and purification. Bottom-up approaches refer to large scale growth by deposition on a substrate. Vapor phase deposition is among the most common techniques and includes chemical vapor deposition (CVD), pulsed laser deposition (PLD), and molecular beam epitaxy (MBE). MBE is a particularly common approach for the synthesis of PTMMCs, since its ultrahigh vacuum environment provides a platform for wafer-scale deposition of highly crystalline and high-purity films. MBE growth aims to be layer-by-layer, so theoretically, any desired number of layers can be deposited. Regardless of the deposition method, the weak binding between the film and the substrate in vdW epitaxy can alleviate lattice mismatch concerns, thus expanding the range of substrate choices.

Research on the epitaxial growth of GaSe is extensive; crystalline GaSe films with thicknesses ranging from a single layer to hundreds of nanometers have been grown on various substrates, including n-type Si,[80] sliced mica,[30] sapphire,[29,80,84-86] GaAs,[86,87] and GaN.[84] Vapor mass transport was the first bottom-up method for growing large-area GaSe nanosheets. However, the growth rate proved too fast for ultrathin film growth.[88] Due to the presence of multiple phases in the Ga-Se phase diagram, it is necessary to precisely control the Se:Ga flux ratio to obtain phase-pure films. Due to the poor wettability of Ga atoms on sapphire, Yu et al.[86] proposed a three-step growth method to improve the quality of atomically thin GaSe films ($\leqslant$ 3 layers). In addition, Shiffa et al.[55] obtained the pure γ'-GaSe polytype on a sapphire substrate using MBE. This is the first experimental report of the synthesis of this polytype using any synthesis method.

Single-crystal α-GaTe films have been synthesized on Si(111) substrates through a two-step MBE growth method based on the phase transition of β-GaTe. Zallo et al.[89] initially grew metastable β-GaTe films at 375°C and induced the phase transition by rapidly heating the films to 550°C for 30 min. The resulting films have thickness ranging from 2 to 18 layers with no grain boundaries and strong photoluminescence signals. Interestingly, the unstable hexagonal β-GaTe is claimed to be the stable phase in the few-layer limit,[59] though the critical thickness at which this transition occurs remains uncertain. Bae et al.[90] grew wafer-scale GaTe films on GaAs substrates via MBE and observed thickness-dependent crystallinity, with hexagonal GaTe at the beginning of growth that transitioned to monoclinic GaTe when the thickness exceeded ~90 nm. Yuan et al.[91] fabricated p-n junctions by growing GaTe multilayers on Si using MBE. The resulting GaTe films exhibited a hole mobility of 28.4 $cm^2V^{-1}s^{-1}$ and a carrier density of $2.5\times10^{18}$ $cm^{-3}$ at room temperature. Vertical Bridgman technique synthesis of α-GaTe crystals followed by mechanical exfoliation is also a common choice for 2D α-GaTe flake fabrication.[31,32,48,61,64,72] The synthesis of α-GaTe typically requires high-temperature annealing to completely transform the metastable hexagonal β-phase into the stable monoclinic α-phase and the injection of Ar gas to promote the formation of GaTe single crystals.[32,92] β-GaTe will always quickly transform to monoclinic α-GaTe after synthesis.[60,62,63]

Okamoto et al.[93] reported the MBE growth of GaS on GaAs substrates using a single tertiarybutyl-gallium sulfide cubane precursor. High growth temperatures of ~500°C yielded epitaxial GaS films with an optical band gap of 2.7 eV, while lower growth temperatures of ~350°C only produced amorphous GaS films with a band gap of 2.2 eV. Highly crystalline and large-area single-/few-layer GaS films have been obtained through micromechanical cleavage using tape.[73] Liquid exfoliation by sonicating layered GaS powders in solvents can produce size-selected nanosheets via controlled centrifugation.[94] Among the bottom-up synthesis methods, the most popular approach for growing GaS crystals is CVD. Wang *et al.*[95] reported CVD growth of GaS with thicknesses ranging from 1 to 15 nm and uniform coverage over an area of 0.7 cm$^2$ by using $Ga_2S_3$ as source material and $H_2$ as carrier gas.

**2.4 Applications**

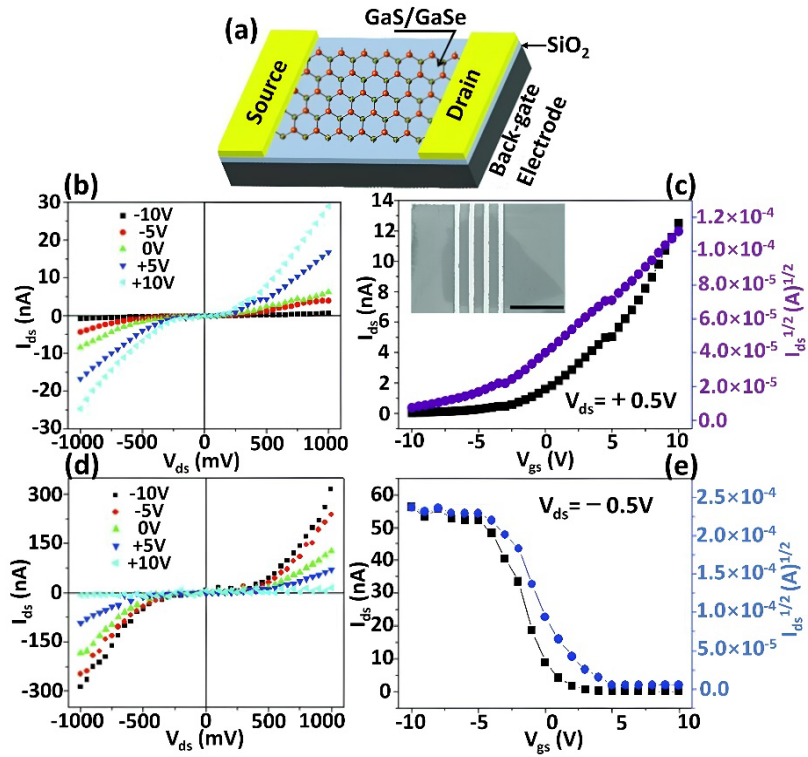

**Figure 6.** (a) A schematic of a typical bottom-gate field effect transistor (FET), based on single sheet of GaS and GaSe as a channel. Output characteristics of single sheet of (b) GaS and (d) GaSe at room temperature in dark. Transfer characteristics of (c) GaS and (e) GaSe single sheet-based FETs. $I_{ds}$, $V_{ds}$, and $V_{gs}$, indicate drain current, drain-source voltage, and gate voltage, respectively. Reprinted with permission from [73]. Copyright 2012 John Wiley and Sons.

Field effect transistors (FETs) consist of a channel connecting the source and drain electrodes, a dielectric barrier separating the gate electrodes from the channel, and the gate electrode itself. The on/off state of the channel current is controlled by changing the gate voltage. Layered GaX have been widely explored as channel materials. Figure 6a displays a typical bottom-gate FET based on single-layer GaS or GaSe with characteristics shown in Figure 6b-e, where the field-effect differential mobilities are 0.1 cm$^2$V$^{-1}$s$^{-1}$ (n-type, for GaS-based FET) and 0.6 cm$^2$V$^{-1}$s$^{-1}$ (p-type, for GaSe-based FET), with on/off current ratios of 10$^4$ (for GaS-based FET) and 10$^5$ (for GaSe-based FET), respectively, which are comparable to the performance of bottom-gate MoS$_2$-based FETs. The 4-layer α-GaTe-based FET exhibits an on/off ratio of 10$^5$ and a hole mobility of 4.6 cm$^2$V$^{-1}$s$^{-1}$,[32] surpassing the performance of FETs based on GaS, GaSe, or MoS$_2$. Furthermore, employing high-K dielectric materials as the top gate to construct top-gate FETs can significantly improve carrier mobility and on/off ratios by reducing the trap/impurity states at the gate surface.

Photodetectors are devices that convert optical energy into electrical signals based on the photoelectric effect. Photo-responsivity, response time, and photo-detectivity are three common performance metrics for evaluating photodetectors. 2D materials are especially exciting as photodetector materials since they are compact, lightweight, and flexible. The first GaS nanosheet-based photodetectors on a rigid SiO$_2$/Si substrate gave a photo-responsivity of 4.2 AW$^{-1}$ at 254 nm.[52] Then an improved device based on few-layer GaS achieved a photo-responsivity of 64.43 AW$^{-1}$ with a response time of 10 ms in an NH$_3$ environment.[81] GaS-based photodetectors fabricated on flexible polyethylene terephthalate substrates are not only highly responsive (19.2 AW$^{-1}$ at 254 nm)[52] but also show ultrahigh specific photo-detectivity of 10$^{12}$–10$^{13}$ Jones for visible light.[51]

Conventional ultrathin GaSe-based photodetectors, when illuminated with 254 nm light, have a response time of 20 ms, a responsivity of 2.8 AW$^{-1}$, and an external quantum efficiency (EQE) of 1367%.[62] The slow response, compared to traditional metal-semiconductor-metal photodetectors with response times around 10$^{-3}$–10$^{-6}$ ms, is ascribed to defects and dangling bonds at the SiO$_2$/GaSe interface. However, the strong photo-responsivity and high EQE make GaSe-based photodetectors attractive. Cao *et al*.[96] further increased the GaSe photo-responsivity to 5000 AW$^{-1}$ by changing the top contact to a bottom contact. They also reduced the response time to 270 μs by replacing Ti/Au electrodes with single-layer graphene. Additionally, Yuan et al.[80] grew GaSe flakes on n-type Si substrates using MBE, forming a p-n heterojunction to fabricate photodiodes with stable rectifying performance and high photo-responsivity. The recently-reported γ'-GaSe with a centrosymmetric unit cell shows resonant ultraviolet (UV) absorption and optical anisotropy, making it effective for deep UV (200–280 nm) sensing.[55] GaSe can not only serve as a photodetector for visible and UV light, but also as an excellent material for nonlinear optical devices due to its large nonlinear optical coefficient of 54 pmV$^{-1}$.[97] Finally, GaTe nanosheet-based photodetectors show high sensitivity and high photo-responsivity in the 254–700 nm range, reaching a peak value of 274.1 AW$^{-1}$ at 490 nm with a comparable specific photo-detectivity of $10^{12}$ Jones.[32,61,72] Compared to GaS and GaSe, the relatively narrow direct bandgap of GaTe greatly enhances photon absorption, making GaTe more suitable for the fabrication of photodetectors.

GaX semiconductors are also at the forefront of many other applications. For example, they are ideal materials for gas sensors because of their high sensitivity to $H_2$ and other gases. They are also candidates for solar cells due to their high optical absorption coefficient. GaS exhibits good transmission in the visible spectrum, which is useful for applications in filters and lenses. GaSe is transparent and has nonlinear optical behaviors in the IR region, thus attracting attention for the development of IR detectors. GaTe, with its high refractive index and good transmission in the UV spectrum, has the potential for UV lens and high-power optical switch applications.

## 3. In Monochalcogenides

### 3.1 Crystal Structure and Band Structure

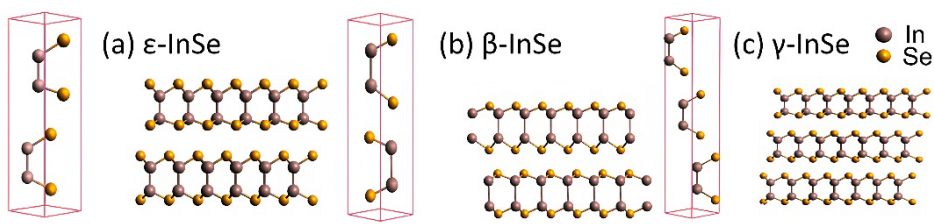

**Figure 7.** Crystal structures of common InSe polytypes. Unit cell (left) and side view (right) of (a) ε-, (b) β-, and (c) γ-InSe. Reprinted with permission from [33]. Copyright 2020 Elsevier.

Since InS[6] and InTe[5] are not layered materials, the focus of this section will be exclusively on InSe. InSe shares almost identical unit cell and stacking configurations with GaSe, but its quadruple layers have a thickness of 8.4 Å. InSe has three common polytypes: ε, β, and γ, as pictured in Figure 7. Table 3 summarizes their crystallographic information. γ-InSe with ABCABC stacking is the most extensively studied polytype.

**Table 3.** Summary of the crystal properties of reported InSe polytypes.[33]

| InSe Type | Stack order | Space group | | Lattice parameters | |
|---|---|---|---|---|---|
| | | Hermann-Mauguin | Schönflies | a=b (Å) | c (Å) |
| ε (2H') | AB | $P\bar{6}m2$ | $D^1_{3h}$ | 4.005 | 16.650 |
| β (2H) | AA' | $P6_3/mmc$ | $D^4_{6h}$ | 4.005 | 16.650 |
| γ (3R) | ABC | $R3m$ | $C^5_{3v}$ | 4.005 | 24.961 |

The band structure of InSe is dependent on the number of layers: increasing the number of layers reduces the bandgap and changes it from an indirect to a direct bandgap semiconductor. Sun et al.[34] simulated the band structures for single-layer and 10-layer β-InSe (Figure 8a,b) and γ-InSe (Figure 8d,e). The thickness dependent bandgap trends for β- and γ-InSe are included in Figure 8c,f. The bandgap of a single-layer InSe is ~2.39 eV irrespective of polytype. For 10-layer β-InSe and γ-InSe, their bandgaps decrease to ~1.24 eV and ~1.18 eV, respectively. From 10-layer to bulk, the bandgaps remain nearly constant.

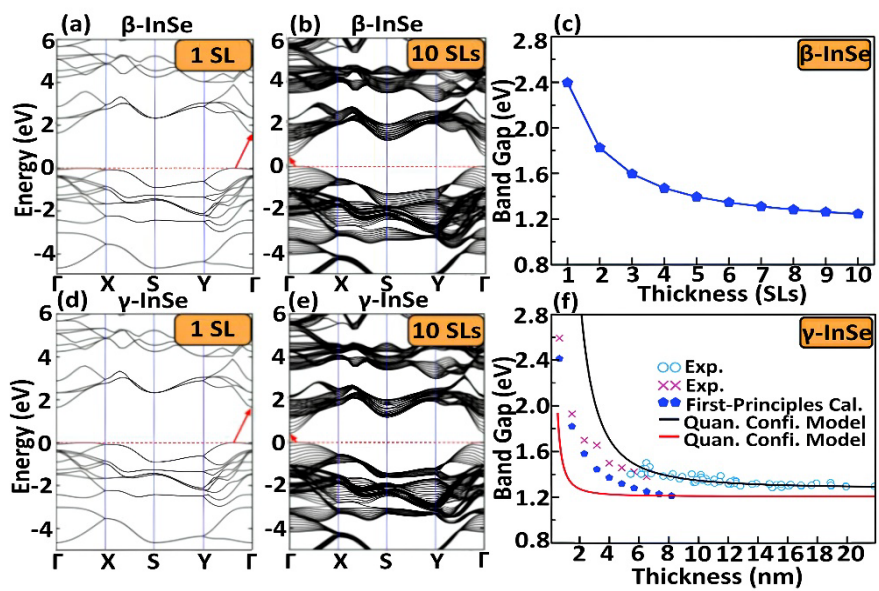

**Figure 8.** Band structures of β-InSe (a) single layer (SL) and (b) 10 SLs, as well as (c) bandgap values for thicknesses from 1–10 SLs. Band structures of γ-InSe (d) SL, and (e) 10 SLs, as well as (f) bandgap values for thicknesses from 1–20 nm. All band diagrams and bandgap values were calculated by GGA-PBE functional. Reprinted with permission from [34]. Copyright 2009 Royal Society of Chemistry.

## 3.2 Electrical and Optical Properties

InSe is typically n-type, and the carriers have a small effective mass and high mobility.[98] Its electrical and optical properties strongly depend on the number of layers. For example, as InSe crystals transition from single-layer to bulk, the electron mobility increases from $10^2$ to $10^3$ cm$^2$V$^{-1}$s$^{-1}$.[99] From 1 to 10 layers, the mobility follows a near-linear trend, and after 10 layers, the carrier mobility saturates. Sucharitakul et al.[49] emphasized the influence of contacts, temperature, and dielectric layers on carrier mobility in InSe multilayers. Under ideal conditions, carrier mobility in InSe can achieve ~$10^3$ cm$^2$V$^{-1}$s$^{-1}$ at room temperature (exceeding $2\times10^3$ cm$^2$V$^{-1}$s$^{-1}$ at 20 K).[100] Despite the excellent carrier mobility, the presence of Se vacancies in the material can lead to long-term instability and degrade the electrical performance. Therefore, encapsulation and passivation techniques are needed.

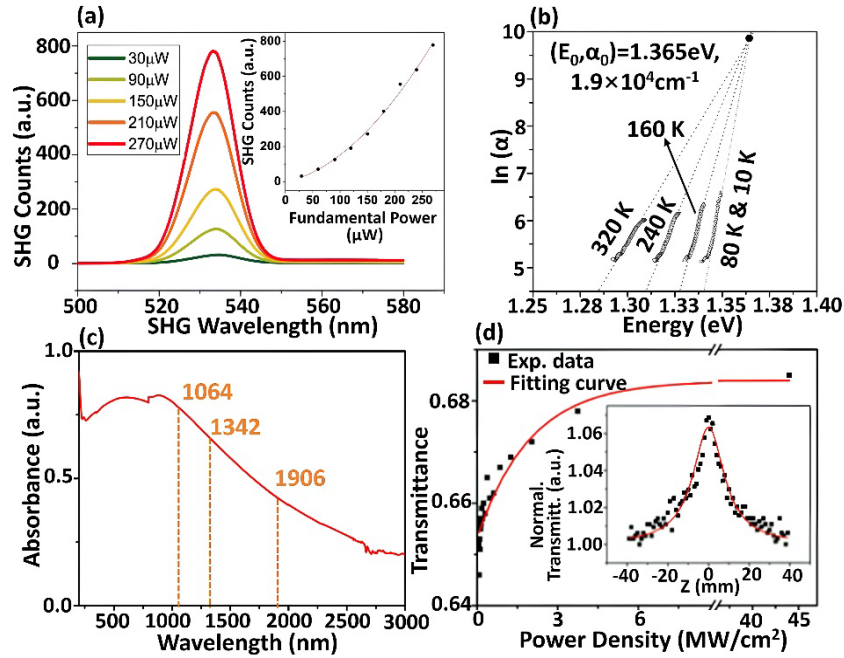

**Figure 9.** Optical properties of InSe. (a) Second-harmonic generation (SHG) spectrum for γ-InSe flakes, where the intensity is scaled in relation to the square of the fundamental beam. Reproduced with permission from [100]. Copyright 2016, Springer Nature. (b) Absorption coefficient (α) vs. incident energy at different temperatures between 10–320 K for bulk InSe. Reprinted with permission from [101]. Copyright 2006 Elsevier. (c) Absorption spectrum of a few-layer β-InSe nanosheet. (d) Transmission spectrum of the InSe nanosheets. (c) and (d) are reprinted with permission from [102]. Copyright 2019 American Chemical Society.

The anisotropic crystal structure of InSe makes it a good material for second-harmonic generation (SHG).[103] In γ-InSe flakes, the SHG intensity increases quadratically with the beam intensity, as shown in Figure 9a. The SHG intensity is a quadratic function of the excitation power, agreeing well with electric dipole theory.[104] Ateş et al.[101] reported that bulk InSe exhibits an absorption coefficient that varies linearly with incident energy in the temperature range of 10–320 K, as shown in Figure 9b. Compared to bulk InSe, the room-temperature absorption spectrum of few-layer InSe nanosheets shows an enhancement in absorption intensity with the incident wavelength ranging from 200 to 800 nm, as depicted in Figure 9c. However, beyond 800 nm, the absorption intensity decreases.[102] The saturation of transmittance with increasing power density, as shown in Figure 9d, further confirms the nonlinear absorption behavior. Even when the incident energy is lower than the bandgap of InSe, defects and edge effects can lead to highly saturated light absorption.[102] In addition, InSe exhibits nonlinear optical refraction, as Yüksek et al.[105] claimed that the refractive index changes non-linearly as the thickness of InSe film increases, transitioning from negative to positive values under ns- or ps-pulse excitation.

**3.3 Synthesis**

Among bottom-up approaches, chemical/physical vapor phase deposition and chemical solution synthesis are common choices for generating wafer-scale high-quality layered InSe films. Common techniques include CVD,[106] MBE,[80,107-110] atomic layer deposition,[111] and PLD.[104,112] InSe nanosheets produced by chemical reactions in solvents rely on precise control of reaction temperature, duration, and precursor molar ratios.[113,114] Common choices for substrates for the growth of InSe films include GaAs(001), graphene, and Si.[80,107-110,115] MBE synthesis of InSe can generally be performed either using a compound $In_2Se_3$ source in a Se-deficient environment,[108,109] or by using separate elemental source materials. In the latter method, the Se:In flux ratio always needs to be precisely controlled to achieve stoichiometric InSe and avoid the formation of unwanted phases such as $In_2Se_3$ and $In_3Se_4$.[115] To date, the most mature method for obtaining 2D InSe nanosheets is still mechanical exfoliation.[116] Bulk InSe crystals are usually prepared through Bridgeman-Stockbarger synthesis.[117] The 2D InSe nanosheets produced using this process are generally high quality, but it is difficult to obtain large-area, high-purity samples with precise control over the number of layers. Liquid phase exfoliation has been introduced due to its low cost and high output,[118] with the quality of the resulting nanosheets depending on ultrasonic power, solvent polarity, and centrifugation rate.[69] N-methyl-2-pyrrolidone, 2-propanol, and ethanol are all effective solvents for producing InSe nanosheets.[70,119] However, this method still faces challenges related to the difficulty in controlling the flake size and thickness, as well as defects and impurities in the lattice.[120]

## 3.4 Applications

InSe is a promising candidate as a channel material for FETs.[121] Early back-gate FETs with SiO$_2$/Si substrates and few-layer InSe as the channel material showed low field-effect electron mobility of ~0.1 cm$^2$V$^{-1}$s$^{-1}$ and unsatisfactory on/off ratio of ~10$^4$ due to interfacial Coulomb impurities, surface roughness, and surface polar phonon scattering from adjacent dielectrics.[122] In addition, air-exposure deteriorates InSe and reduces field-effect mobility.[123] To address these issues, Feng et al.[122] used a polymethyl methacrylate (PMMA)/SiO$_2$ bilayer as the bottom dielectric for back-gate InSe FETs, as shown in Figure 10a, improving the field-effect mobility to 1055 cm$^2$V$^{-1}$s$^{-1}$. Hexagonal boron nitride (h-BN) encapsulation, as illustrated in Figure 10b, was proven to effectively suppress the degradation of InSe in oxygen- and moisture-free environments, yielding electron mobilities ranging from 30 to 120 cm$^2$V$^{-1}$s$^{-1}$ (the control group showed ~1 cm$^2$V$^{-1}$s$^{-1}$ for an isolated InSe layer).[124] The performance has remained stable even after 15 days of exposure to air.[125] Lastly, Huang et al.[126] achieved a non-rectifying barrier by forming Ohmic contacts through In deposition on the entire flake and contact area.

InSe also serves as the ferroelectric channel material for ferroelectric FETs (Fe-FETs). Ferroelectric materials are characterized by a switchable spontaneous polarization under an external electric field, which are useful for nonvolatile memory transistors, microwave devices, and radio frequency devices. The ferroelectricity of InSe was first reported by Hu et al.[127] in mechanically exfoliated β-InSe nanoflakes. Later, they determined that the mechanism for ferroelectricity in InSe is relative layer sliding for in-plane polarization and atom displacement under electric field for out-of-plane polarization.[128] Based on this, they manufactured bottom-gate Fe-FETs that exhibited a high on/off ratio of ~10$^4$. An Fe-FET based on CuInP$_2$S$_6$/h-BN/InSe was developed by Singh et al.,[129] which showed outstanding performance as nonvolatile memory with a large memory window (4.6 V at a voltage sweep of 5 V), high current on/off ratio (> $10^4$), high endurance, and long data retention (> $10^4$ s).

Given the outstanding electronic and optical properties of layered InSe, it has enormous potential to advance photodetector applications. Figure 10c shows an optical image of a photodetector based on Gr-InSe-Gr with graphene multilayers as electrodes.[130] Ohmic contacts are formed between InSe and graphene near the interface, enhancing electron injection compared to traditional Schottky contacts. Consequently, this photodetector displayed an EQE of up to 5000 under illumination at 633 nm. Furthermore, doping engineering is widely employed to tailor the performance of 2D layered semiconductors. For instance, Chang et al.[131] exploited surface (~2 nm deep) oxygen doping to enhance the performance of InSe-based photodetectors. Hao et al.[130] designed an S-doped InSe photodetector, Gr/InSe$_{0.9}$S$_{0.1}$/Gr, which achieved a maximum responsivity of up to $4.9 \times 10^6$ mAW$^{-1}$. However, similar to GaSe, InSe degrades under the influence of air and light, making encapsulation or capping necessary.[125]

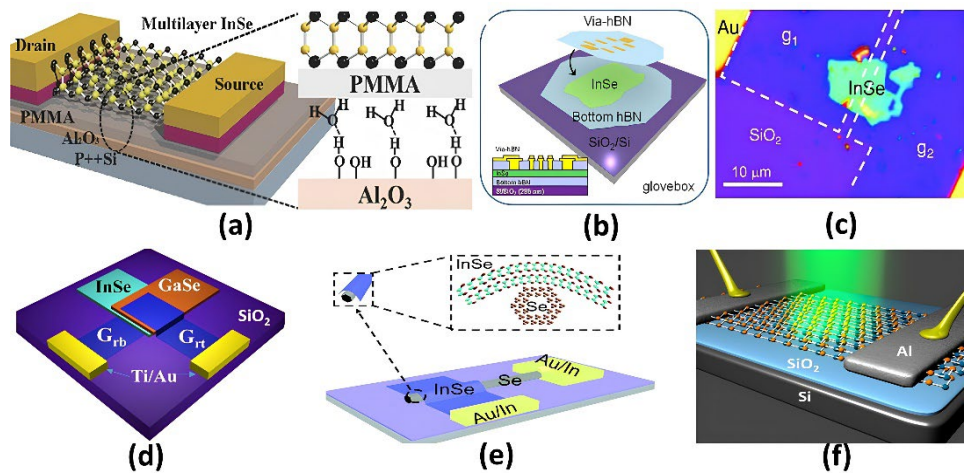

**Figure 10.** Examples of layered InSe nanosheet applications. (a) a back-gate InSe FET with PMMA/SiO$_2$ bilayer dielectric, reprinted with permission from [122], copyright 2014 John Wiley and Sons; (b) an h-BN encapsulated InSe FET, reprinted with permission from [125], copyright 2016 Royal Society of Chemistry. (c) Optical image of a Gr-InSe-Gr based photodetector, where g$_1$ and g$_2$ indicate two graphene areas serving as the source and drain to InSe channel of the gate structure. Reprinted with permission from [130]. Copyright 2015 John Wiley and Sons. Schematics of (d) an InSe/GaSe p-n photodiode, reprinted with permission from [133], copyright 2019 American Chemical Society; (e) a Se-InSe heterojunction photodiode, reprinted with permission from [134], copyright 1990 IOP Publishing, Ltd.; (f) an InSe nanosheet-based avalanche photodetector with Al contacts, reprinted with permission from [135], copyright 2015 American Chemical Society.

Self-powered photodetectors, which do not rely on external power sources for operation, are superior to conventional photodetectors, and InSe heterojunction photodiodes demonstrate a successful realization of this concept. Yan et al.[133] assembled a photodiode based on an InSe/GaSe p-n heterojunction, as shown in Figure 10d. Under illumination at 410 nm, it achieved a responsivity of 21 mAW$^{-1}$ and a rapid response time of 5.97 µs. Similar cases exist using InSe/GaTe and Gr/InSe/MoS$_2$ heterojunctions.[136] Materials such as elemental Se or Te have also been used as the p-type component in p-n heterojunctions with InSe, as shown in Figure 10e. Moreover, Lei et al.[135] fabricated an InSe nanosheet-based avalanche photodetector, where InSe and Al electrodes form Schottky contacts, as depicted in Figure 10f. It exhibited outstanding EQE of 1110% with an avalanche gain of 152, and rapid response within 60 µs at a bias voltage of 70 V.

## 4. Ge Monochalcogenides

### 4.1 Crystal Structure and Band Structure

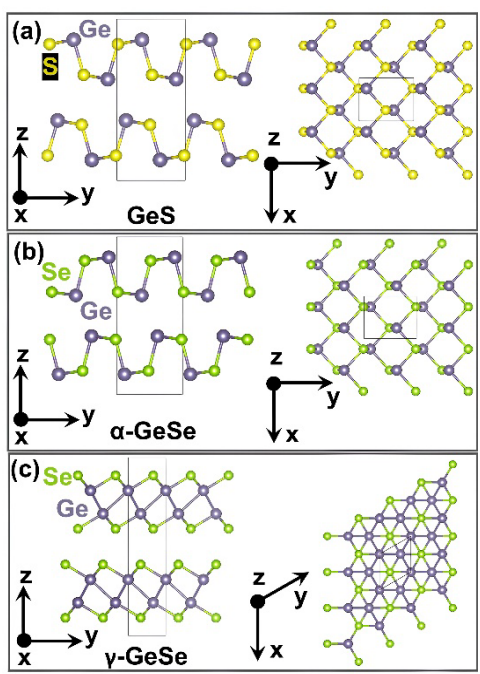

**Figure 11.** Sketches of the crystal structures of (a) orthorhombic *Pnma* GeS with a=3.60 Å, b=4.52 Å, and c=10.78 Å; (b) orthorhombic *Pnma* α-GeSe with a=3.83 Å, b=4.39 Å, and c=10.84 Å; (c) hexagonal *P6₃mc* γ-GeSe with a=b=3.73 Å, c=15.40 Å. Only one single layer is sketched in the displayed z-zone axis for each material.

The crystal structures of the experimentally realized vdW layered GeX' under ambient pressures and temperatures are sketched in Figure 11.[137-147] The puckered crystal structure of GeS and α-GeSe is close to black phosphorous (BP). Both were classified as vdW materials, although their interlayer binding energies range on the upper end of easily exfoliable vdW materials suggesting a small component of ionic bonding across individual single layers.[140,141,148,149] α-GeSe was reported to transform subsequently into a *R3m*, *Fm$\bar{3}$m*, and eventually the β-GeSe phase – also BP-like, but with a proportionally much larger lattice constant b– under increasing pressures.[138,150,151] γ-GeSe is the youngest known and realized 2D layered vdW phase of the GeX. Its hexagonal crystal is very similar to γ-InSe and therefore much different to the puckered BP-like structures of GeS, and α-GeSe.[139] Although more 2D layered vdW phases were identified by computer calculations for GeX (The Materials Project and Materials Cloud), they were found thermodynamically unstable and/or have not been successfully synthesized today.[140-147,152,153] The thermodynamically stable monochalcogenide GeTe, on the other hand, forms 3D MVB structures that are not part of the layered PTMMC family.[6,25,26,142-147,154-156]

Density functional theory (DFT) calculations estimate single-layer GeS to be an indirect bandgap semiconductor with a band gap of ~2.3 eV and α-GeSe to have a smaller direct bandgap of ~1.5 eV.[39] These predictions were experimentally confirmed by multiple groups: GeS and α-GeSe have been shown to be p-type semiconductors with band gaps around 1.55–1.71 eV and 1.1–1.53 eV, respectively.[42,43,157,158] The ternary compound $GeSe_{1-x}S_x$ showed excellent band gap tuning from 1.14 to 1.71 eV as the composition varied from x=0 to x=1.[158] Band diagrams for the single layers of GeS and α-GeSe are depicted in Figure 12a,b, respectively. Bulk GeS remains an indirect semiconductor with an established band gap of 1.56 eV, whereas reported direct and indirect band gaps of bulk α-GeSe range from 0.8–1.54 eV at room temperature.[42,43,138,159-162] As shown in Figure 12c, γ-GeSe is an indirect semiconductor in the single layer limit with a calculated energy gap in the 0.99–2.96 eV range.[163-165] Bulk γ-GeSe shows semimetallic properties owing to its high p-doping level and very small indirect band gap at or below 0.36 eV.[139,163-166]

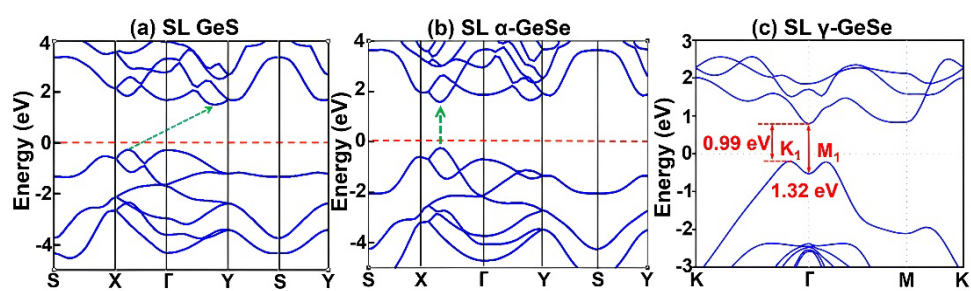

**Figure 12.** Electronic band structures of single-layer (SL) (a) GeS, and (b) α-GeSe, reprinted with permission from [39], copyright 2017 American Chemical Society; (c) γ-GeSe, reprinted with permission from [164], copyright 2014 IOP Publishing, Ltd.

### 4.2 Electronic and Optical Properties

GeS is predicted by DFT calculations to show absorption in the visible and UV

spectra.[44] The direct optical transitions in GeS are highly anisotropic along the two in-plane directions due to the puckered nature of its crystal structure.[159] For single-layer GeS, the electron and hole mobilities were predicted to reach 2750 and 220 $cm^2V^{-1}s^{-1}$, respectively.[167] α-GeSe is isostructural and isoelectronic to BP and has strong in-plane anisotropic properties.[138,166] A carrier mobility of 128.6 $cm^2V^{-1}s^{-1}$ [161] was found for α-GeSe along with an room-temperature resistivity of 0.82 Ω cm.[138] The resistivity increases dramatically with decreasing temperature below ~100 K, but decreases slightly with decreasing temperatures above ~200 K due to the high mobility of carriers.[138] α-GeSe also exhibits strong light absorption with an absorption coefficient exceeding $10^4$ $cm^{-1}$ in the visible range[43] and an ultrahigh photo-responsivity of $1.6×10^5$ $AW^{-1}$.[161] The newly discovered polytype, γ-GeSe, possesses an exceptionally high hole carrier concentration in the range of $5×10^{21}$ $cm^{-3}$ with hole mobilities reaching 12.2 $cm^2V^{-1}s^{-1}$ at room temperature,[139,166] leading to excellent electrical conduction with a low resistivity of $1.35×10^{-4}$ Ω cm. The metal-like behavior originates from ~5% Ge vacancies in the unconventional configuration in γ-GeSe.[166] The resistivity of γ-GeSe increases with temperatures between 50–300 K, indicating that electron-photon scattering dominates. First principle calculations predicted further superior light absorption in the order of $10^6$ $cm^{-1}$ in the UV and visible spectra for single-layer γ-GeSe topping that of α-GeSe by two orders of magnitude. Moreover, weak antilocalization resulted from spin-orbit coupling (SOC) and the spontaneous out-of-plane polarization as well as ferroelectricity caused by the non-centrosymmetric atomic structure were also observed in the synthesized γ-GeSe.[163,166] Unlike α- and γ-GeSe, the resistivity of β-GeSe shows temperature-independent characteristics, suggesting that it could be dominated by defect-state conduction, and is 0.55 Ω cm at room temperature.[138]

### 4.3 Synthesis

The biggest challenge hindering the study of Ge-related chalcogenides is the large-scale manufacturing of these materials in the crystalline phase. GeS and α-GeSe have been synthesized by high-energy mechanical milling, CVD, laser irradiation, Bridgman, chemical vapor transport techniques followed by mechanical exfoliation, physical vapor deposition (PVD), colloidal synthesis, plasma discharge of precursors, electrodeposition, and chemical reduction. These processes yield various nanocomposites and nanosheets.[42,137,168,169] GeS nanowires have also been synthesized using a vapor-liquid-solid growth process, but unfortunately exhibited metal contamination from the metallic liquid catalyst.[169] α-GeSe nanobelt formation has been reported using a colloidal synthesis approach.[170] Unfortunately, exfoliated flakes and crystalline nanosheets reach only 10–200 μm. Recently, Zhang et al.[169] have pushed the lateral size of polycrystalline GeS clusters grown in PVD-grown films up to square centimeter-sized patches with a thickness of 100 nm on $SiO_2$/Si substrates.

Typically, solution-synthesized GeS and α-GeSe nanosheets perform better in photodetection compared to their vapor deposition and mechanical exfoliation counterparts.[157] However, reports on solution-based synthesized GeX' are rare due to the easy oxidation of the intermediate Ge (II) to Ge (IV) in the precursor solution, facilitating $GeS_2$ and $GeSe_2$ formation over their monochalcogenide counterparts.[158] Hu et al.[158] have presented an alteration to the solution-based fabrication from Ge (II) to achieve α-GeSe and GeS films by introducing hypophosphorous acid as a suitable reducing agent and strong acid to generate Ge (II) from $GeO_2$ powders. The solution-based fabrication process proved to be a reliable method for α-GeSe, GeS, and ternary $GeSe_{1-x}S_x$ films. However, the resulting films were polycrystalline and contained many grain boundaries, limiting the electrical and optoelectronic performance of the films. To increase the crystallinity and the lateral size of α-GeSe grains in thin films, Chen et al.[43] developed a thermal evaporation process under high vacuum conditions starting from GeSe powders. By sandwiching two GeSe films on $TiO_2$ coated soda lime glass substrates together and annealing them post growth in a quartz tube flowing Ar gas, they realized single-oriented GeSe(100) films. Crystalline nanostructures of γ-GeSe have very recently been experimentally realized by CVD exclusively.[139,166] Using polycrystalline α-GeSe powder as deposition source in CVD, the key to obtain γ-GeSe formation turned out to be a nm-thin Au layer deposited on the substrate prior to CVD growth acting as a catalyst in a vapor-liquid-solid growth mode that produced highly tapered dagger-shaped γ-GeSe flakes of about 20 μm length, and 40 nm thickness.[139] On the other hand, the key to synthesizing single-crystal β-GeSe lies in high-temperature (e.g., 1200 °C) and high-pressure (e.g., 6 GPa) environments.[138] Despite their interesting properties, no reports on the MBE growth of GeS and GeSe exist to date.[149]

### 4.4 Applications

Based on their bandgaps and band edges, both GeS and α-GeSe have attracted much attention for their electrochemical performance. Calculations predict high electrocatalysis potential for both composites,[39,171] but bulk GeS fell short for both hydrogen evolution reaction (HER) and oxygen reduction reaction (ORR).[168] α-GeSe was predicted and experimentally confirmed to perform better in HER and ORR than GeS due to better band edge locations and higher photo absorption coefficient in the visible range. Ge vacancies increase this potential as these vacancies decrease the antibonding state occupation number and generate more available bonding states with hydrogen atoms. The hydrogen bonded material becomes metallic, which is of further advantage due to the increased conductivity.[43,171] The inherent electroactivity of GeS can be increased and its HER activity is boosted by an electrochemical pre-treatment, which removes the passivating oxide surface layer.[168] GeS nanocomposites furthermore show high potential for full-cell lithium-ion batteries due to the large capacity, high first coulombic efficiency, safe operating voltage, and excellent rate performance.[137]

GeS and α-GeSe nanosheets have both shown great potential in photovoltaic (PV) applications with almost perfect bandgap energies, excellent optoelectronic properties, earth-abundancy, and non-toxicity. α-GeSe thin-film PV devices have recently reached an efficiency of 5.2% with air stability.[158] GeS PV devices are advantageous for indoor use due to the wider band gap of GeS, which matches well with the emission spectra of commonly used indoor light sources. In the ternary alloy GeSe$_{1-x}$S$_x$, the tunable band gap between 1.1–1.71 eV enables the fabrication of optimal-bandgap single- and multi-junction PV devices.

Preliminary GeS PV devices on glass showed a power conversion efficiency of 0.8%, falling far short of the theoretical limit of 30%, indicating a need to find high-quality crystalline substrates for Ge monochalcogenide PV devices.[158] α-GeSe single-junction PV devices have reached 1.48% power conversion efficiencies with an open-circuit voltage of 240 mV.[162] Chen et al.[43] improved the open-circuit voltage to 450 mV (average value of 340 mV) by forming better crystalline α-GeSe films through sandwich annealing GeSe films on TiO$_2$ buffer layers. Unfortunately, the corresponding power conversion efficiency of 0.27% is far from satisfactory; this low value is thought to be caused by the thin depletion width and high interface recombination at the TiO$_2$/GeSe junction. In addition to the ideal bandgap for PV applications, α-GeSe shows remarkable air stability and strong light absorption with an absorption coefficient exceeding $10^4$ cm$^{-1}$ in the visible range, which equals full light absorption within a 1 μm thick layer.[43] In addition, GeS and α-GeSe nanosheets are popular candidates in photodetector applications with high photosensitivity as well as fast and stable photo response.[157]

Thermoelectric devices are based on a reversible energy conversion between heat and electricity. Their performance is usually evaluated by a thermoelectric figure-of-merit (ZT) that depends on the Seebeck coefficient (S), electrical conductivity (σ), thermal conductivity (κ), and temperature (T), following the equation ZT=S$^2$σT/κ. High-efficiency thermoelectric materials thus require a low thermal conductivity and a high electrical conductivity. Although γ-GeSe was shown to possess a low Seebeck coefficient,[166] its low lattice thermal conductivities (~1.72 Wm$^{-1}$K$^{-1}$ for bulk and <1 Wm$^{-1}$K$^{-1}$ for single-layer at 900 K) and high electrical conductivity yielded high ZT values (0.8 for bulk and 2.8 for single-layer at 900 K) for thermoelectric applications.[165] The significantly enhanced ZT values of γ-GeSe with decreasing dimensionality motivates the development of 2D thermoelectric devices. The highest ZT value of 3.8 was calculated for 2-layer γ-GeSe at 800 K.[165] Besides the interesting thermoelectric properties, γ-GeSe is predicted to show superior light absorption reaching $10^6$ cm$^{-1}$ in the single-layer limit.[164] Combined with its band diagram rich in valleys at band edges that enable high carrier concentrations and low direct electron-hole recombination rates elevates γ-GeSe to a potential candidate for high-performing optoelectronic applications.[164] A surprisingly low and nearly isotropic Young's modulus of 86.59 Nm$^{-1}$ (amounting to only ¼ of the graphene value) and a Poisson's ratio of 0.26 is furthermore predicted for single layer γ-GeSe, making it an intriguing material for flexible electronics and strain engineering.[164]

## 5. Sn Monochalcogenides

### 5.1 Crystal Structure and Electronic Structure

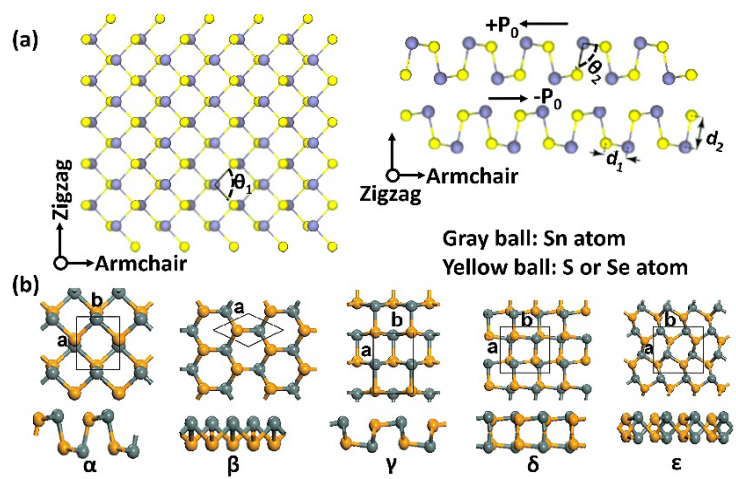

**Figure 13.** (a) Top (left sketch) and side (right sketch) views of the orthorhombic crystal structure of bulk α-SnX'. $\theta_1$=96.8° for SnS, while $\theta_1$=96° for SnSe. $\theta_2$=89° for both SnS and SnSe. Reprinted with permission from [172]. Copyright 2020 Elsevier. (b) Top (top panel) and side (bottom panel) views of the crystal structures for single-layer SnX' with different phases. Reprinted with permission from [35]. Copyright 2009 Royal Society of Chemistry.

Crystalline SnS and SnSe belong to the layered vdW family, while SnTe is a 3D MVB material.[25,26,36,173] Bulk SnX' have thermodynamically stable orthorhombic crystal structures in either the α-phase with a *Pnma* space group or the β-phase with a *Cmcm* space group. Table 4 summarizes the lattice parameters for bulk SnX'. The phase adopted by SnX' depends largely on temperature, with the phase transition temperature (from α- to β-phase) being 873 K for SnS[36] and 800 K for SnSe.[174] At room temperature, both SnS and SnSe exist in the α phase. Figure 13a depicts the crystal structure of bulk α-SnX'. Each unit cell contains four Sn atoms and four chalcogen atoms in four mixed Se-X' layers, where four atomic layers form a puckered quadruple layer, and two quadruple layers are bonded by vdW forces at a distance ~4 Å.[36,173] Within a unit cell, each Sn atom is covalently bonded to three adjacent chalcogen atoms, and a lone pair left by $Sn^{2+}$ exhibits a distortion in the orthorhombic structure. Thus, compared to the perfect 90° bond angle in BP structure, slight deviations in the cation-anion bond angles ($\theta_1$ and $\theta_2$) are expected for α-SnX', as demonstrated in Figure 13a. This results in a lattice structure without inversion symmetry, which induces a strong in-plane spontaneous ferroelectric polarization in SnX' structures.[175] For the single-layer SnX', the α phase remains the most studied structure, as its lattice parameters closely match those of their bulk counterparts despite belonging to a different space group of *Pmn2₁*.[173] Other polytypes of SnX' single layers have been identified by theoretical studies; their crystal structures and lattice parameters are given in Figure 13b and Table 4, respectively.

**Table 4.** Crystal information about SnX'.[35-38]

| Type | Structure | Space group | | a (Å) (armchair direction) | b (Å) (zigzag direction) | c (Å) (out of plane) |
| --- | --- | --- | --- | --- | --- | --- |
| | | Hermann-Mauguin | Schönflies | | | |
| Bulk | α-SnS | *Pnma* | $D^{16}_{2h}$ | 4.34 | 3.97 | 11.14 |
| | β-SnS | *Cmcm* | $D^{17}_{2h}$ | 4.17 | 11.48 | 4.12 |
| Single-layer | α-SnS | *Pmn2₁* | $C^{7}_{2v}$ | 5.84 | 3.79 | -- |
| | β-SnS | *P3m1* | $C^{1}_{3v}$ | 6.50 | 3.75 | -- |
| | γ-SnS | *Pmn2₁* | $C^{7}_{2v}$ | 5.84 | 3.79 | -- |
| Bulk | α-SnSe | *Pnma* | $D^{16}_{2h}$ | 4.14 | 4.44 | 11.49 |
| | β-SnSe | *Cmcm* | $D^{17}_{2h}$ | 4.31 | 11.70 | 4.31 |
| **Single-layer** | α-SnSe | *Pmn2₁* | $C^7_{2v}$ | 3.95 | 4.82 | -- |
| | β-SnSe | *P3m1* | $C^1_{3v}$ | 3.78 | 3.78 | -- |
| | γ-SnSe | *Pmn2₁* | $C^7_{2v}$ | 3.78 | 6.11 | -- |
| | δ-SnSe | *P2₁ca* | $C^5_{2v}$ | 6.14 | 6.23 | -- |
| | ε-SnSe | *P2₁ca* | $C^5_{2v}$ | 7.10 | 6.60 | -- |

Most studies of the electronic band structures of SnX' have been done using DFT.[35-37,173,176-179] Bulk SnX' have an indirect transition with an energy gap of 1 eV for SnSe and 1.24 eV for SnS.[173] These results agree well with experimental values published by Parenteau and Carlone,[180] where the bandgap energy of SnS varies between 1.049–1.076 eV, and that of SnSe varies between 0.898–0.903 eV at room temperature. They further determined that the energy gap for direct inter-band transitions in bulk SnX' (1.296 eV for SnS and 1.047–1.238 eV for SnSe) is only slightly higher than that for indirect transitions.

As the layer thickness decreases, a substantial increase in the bandgap energy can be observed, which is primarily attributed to the stronger Coulomb interactions and smaller dielectric screening effects in the lower dimensional forms of these materials.[177-180] The stronger Coulomb interaction results in higher exciton binding energies in their single-layer forms. In single-layer SnS, exciton binding energies have been estimated to be in the range of 0.5–0.9 eV, far exceeding that in bulk SnS ($\leqslant$ 0.1 eV).[181] The large exciton binding energies in single-layer SnS (~0.5 eV) and SnSe (~0.28 eV) also make them promising for applications in nanoscale optoelectronic devices. The five single-layer polytypes of SnSe have different band structures.[35,176] The calculated band gap energies of α-, β-, and γ-SnS are 0.86 eV, 2.30 eV, and 1.46 eV, respectively, while the calculated band gap energies of α-, β-, γ-, δ-, and ε-SnSe are 1.04–1.67 eV, 2.22–2.82 eV, 1.52–2.09 eV, 1.55–2.12 eV, and 1.50–1.93 eV, respectively. The most interesting polytype is single-layer ε-SnSe, which is the only type with a direct band gap. Moreover, the highly dispersive conduction band minimum and valence band maximum in single-layer ε-SnSe suggest that effective masses of both holes and electrons are very small.[35]

## 5.2 Electronic and Optical Properties

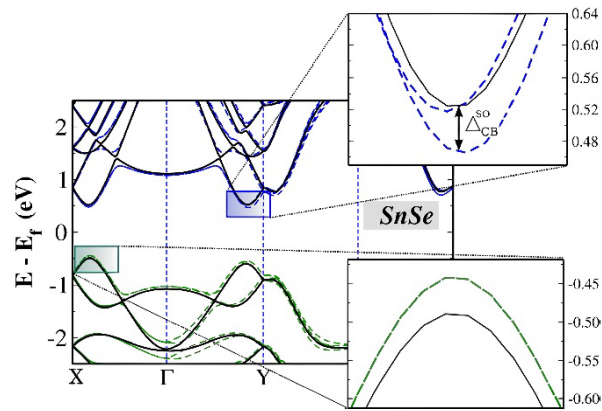

**Figure 14.** Electronic band structure of single-layer SnSe with (green and blue dashed lines) or without (solid black lines) SOC effect. Reproduced with permission from [173]. Copyright 2015, American Physical Society.

SnX' thin films have a strong optical response from the visible to the near-IR and large absorption edges with good electrical conductivity, which is ideal for PV devices. The lack of inversion symmetry in SnX' leads to interesting electronic and optical properties. SOC is a phenomenon where the interaction between the spin of electrons and the orbital motion leads to changes in atomic energy levels.[173,177] Theoretically, this spin-orbit interaction can only be observed in odd layers of SnX' because the inversion symmetry and time-reversal symmetry in even layers and bulk counterparts result in spin degeneracy. Gomes and Carvalho[173] were among the first to report the SOC effect in single-layer SnX', as shown in Figure 14. There are two splits occurring at the local minima of the conduction band valley, along the Γ-X direction and the Γ-Y direction with an energy difference of 38 meV and 52 meV, respectively; the split along Γ-Y defines a new conduction band minimum. Simultaneously, the valence band has an energy split of 14 meV. Similar behaviors have been found in single-layer SnS with a conduction band splitting of 87 meV and a valence band splitting of 8 meV. The spin-orbit splitting in the conduction band valley of single-layer SnX' is higher than that of TMDs, making SnX' promising in the field of spintronics.

Piezoelectricity is also found in single-layer SnX', arising from the lack of inversion symmetry.[37,177,182,183] Multiple research teams have theoretically predicted large piezoelectric coefficients for single-layer SnX'. Compared with other common piezoelectric materials like h-BN and MoS$_2$, the piezoelectric coefficients e$_{11}$ of single-layer SnX' are an order of magnitude larger, while their elastic constants C$_{11}$ and C$_{12}$ are at least two times smaller. These combined factors lead to very large piezoelectric tensor coefficients d$_{11}$ of 144.76 pmV$^{-1}$ in single-layer SnS and 250.58 pmV$^{-1}$ in single-layer SnSe, making them outstanding candidates for piezoelectric devices.

The in-plane anisotropy in the structures of single-layer SnX' leads to different carrier mobilities along the armchair and zigzag directions, in the range of $10^2$–$10^4$ cm$^2$V$^{-1}$s$^{-1}$, which are higher than other vdW PTMMC materials. Layered heterostructure devices composed of single-layer SnX' and other vdW materials can achieve even higher carrier mobilities. For example, a GeSe/SnSe heterostructure displayed a hole mobility of 6.42×10$^4$ cm$^2$V$^{-1}$s$^{-1}$.[184] SnX' layers also possess strong nonlinear optical properties as a result of their structural anisotropy. Large SHG tensor components of up to 10 nmV$^{-1}$ have been predicted in both SnS and SnSe single layers, which are among the largest reported values in 2D materials.[185] Finally, the different polytypes of single-layer SnSe have different absorption, reflection, and refraction properties.[176] The absorption edges of α- and δ-SnSe fall in the near-IR range, whereas those of β-, γ-, and ε-SnSe are in the visible region. In γ-SnSe crystals, visible light is absorbed along the x-axis but transmitted along the z-axis. α-SnSe is polarization-sensitive, with light polarized along the x-axis only absorbed in the energy range of 1.87–2.34 eV.

## 5.3 Synthesis

Unlike most vdW materials, exfoliation is not a common method for obtaining single- or few-layer SnX' compounds because SnS and SnSe both have large exfoliation energies (> 30 meVÅ$^{-1}$).[149] Thus, exploring alternative synthesis methods is crucial. MBE is one of the most successful methods for synthesizing SnSe single layers. Chang et al.[186] used high-purity SnSe granule source materials to grow highly oriented SnSe single layers on graphene substrates using a two-step growth and anneal process. SnSe nanoplates have been produced by optimizing the growth and annealing temperatures.[149,187,188] Maintaining stoichiometry is critical for the growth of SnSe. Chin et al.[187] used individual elemental source materials to study the self-limiting stoichiometry in SnSe and revealed that within a limited range, changing the Se:Sn flux ratio does not influence the film stoichiometry, but only rotates the primary crystallographic orientation from (210) to (200). By replacing MgO(001) substrates with a-plane sapphire substrates, which have rectangular lattice-matched symmetry that breaks the SnSe domain degeneracy, Mortelmans et al.[188] realized SnSe films free of 90° twin defects.

Although large-scale production of SnS single layers using MBE has not been achieved, teams have reported successful synthesis of few-layer SnS. Bao et al.[189] conducted trials to grow 1–7 layers of SnS on mica, MgO(001), Au(111), and graphene substrates. They first evaporated SnS powders and deposited them on the substrate at 200 °C to form a SnS film, which was then annealed to expand the crystal domains. For better control of stoichiometry, Li et al.[190] utilized separate Knudson cells to provide Sn and S fluxes independently and obtained SnS films with thicknesses as low as 18 layers on mica and Nb-doped $SrTiO_3$(100) substrates, respectively. More importantly, the authors found a reversible conversion between SnS and $SnS_2$ crystals via sulfurization.

PVD is another potential option for synthesizing SnS single layers. Higashitarumizu et al.[191] grew SnS single layers on mica substrates using a home-made PVD furnace (Figure 15a) and SnS powders as a source supply. During the growth, $N_2$ gas was introduced and the substrate temperature was controlled as Figure 15b illustrates. Figure 15c shows optical images of multiple circular SnS flakes with a diameter of ~5 μm. The thickness of these flakes can be as low as a single layer (~0.7 nm thick). For single-layer SnSe nanosheets, Jiang et al.[192]

proposed a two-step fabrication method. Initially, bulk SnSe crystals were deposited on $SiO_2/Si$ substrates using a vapor transport deposition system under atmospheric $Ar/H_2$ conditions. Next, the bulk crystals were transferred upstream inside the furnace with an injection of $N_2$ gas at high temperatures. This process yielded large-area rectangular single-layer SnSe flakes with a dimension of 30×50 μm². Liquid-phase exfoliation also enables the large-scale production of ultrathin SnX' nanosheets with thicknesses as low as bilayers.[193]

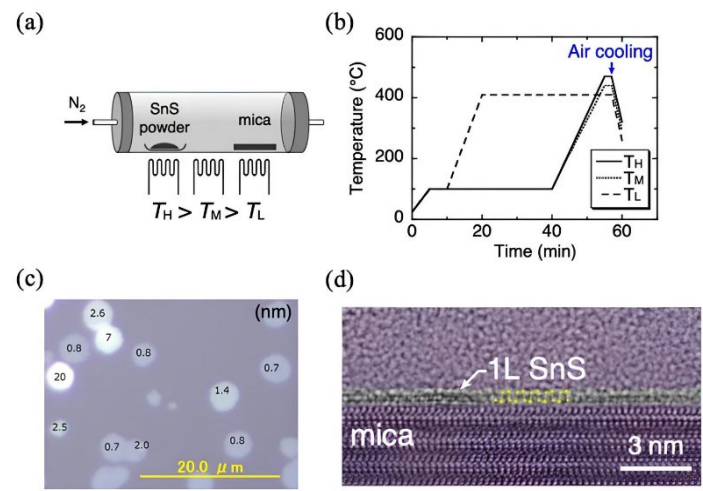

**Figure 15.** (a) Schematic of a three-zone PVD chamber by separately controlling the heaters at high, middle, and low temperatures ($T_H$, $T_M$, and $T_L$). (b) Temperature profiles for SnS PVD growth. (c) An optical image of SnS flakes on mica. The number on each flake indicates the number of SnS layers. (d) A cross-sectional transmission electron microscopy image of a SnS single layer along the armchair direction. Reproduced with permission from [191]. Copyright 2020, Springer Nature.

**5.4 Applications**

SnX' crystals are superior thermoelectric materials with high ZT coefficients, i.e., 1.75–1.88 for SnS and 2.3–2.63 for SnSe at 700 K.[35,38,174,177,179,194-196] It is worth noting that the high ZT coefficient is usually only found at high temperatures where the β-SnSe phase forms, although the Seebeck coefficient of α-SnSe is higher than that of β-SnSe.[197,198] Recently, a ZT of 2.63 along the b-axis for bulk SnSe crystals was reported by Zhao et al. with a ZT of 2.3 along the a-axis.[38] Moreover, C. Zhou et al.[199] reported a p-doped polycrystalline SnSe sample exhibiting an extremely high ZT value (up to 3.1 at 783 K) with an ultralow lattice thermal conductivity of 0.07 Wm$^{-1}$K$^{-1}$. Such exceptional thermoelectric performance is attributed to the ultralow thermal conductivity of bulk SnSe and the anharmonic and anisotropic bonding in its crystal structure. However, the challenge lies in achieving good ZT values at room temperature: the ZT of SnSe at 300 K is less than 0.5 in all directions. Nano-structuring is a common choice to raise room-temperature ZT values.[35,194-196] Wang et al.[196] predicted a high ZT of 3.27 at 300 K along the armchair direction for SnSe single layers, which is seven times higher than its bulk counterpart. The high density of states near the Fermi energy and strong carrier quantum confinement in low-dimensional SnX' materials enhance the power factor (PF=S$^2$σ).[200] Finally, Hu et al.[35] found a relatively low thermal conductivity of 0.156 Wm$^{-1}$K$^{-1}$ and a high ZT of 2.06 at room temperature in a honeycomb-structure single-layer β-SnSe.

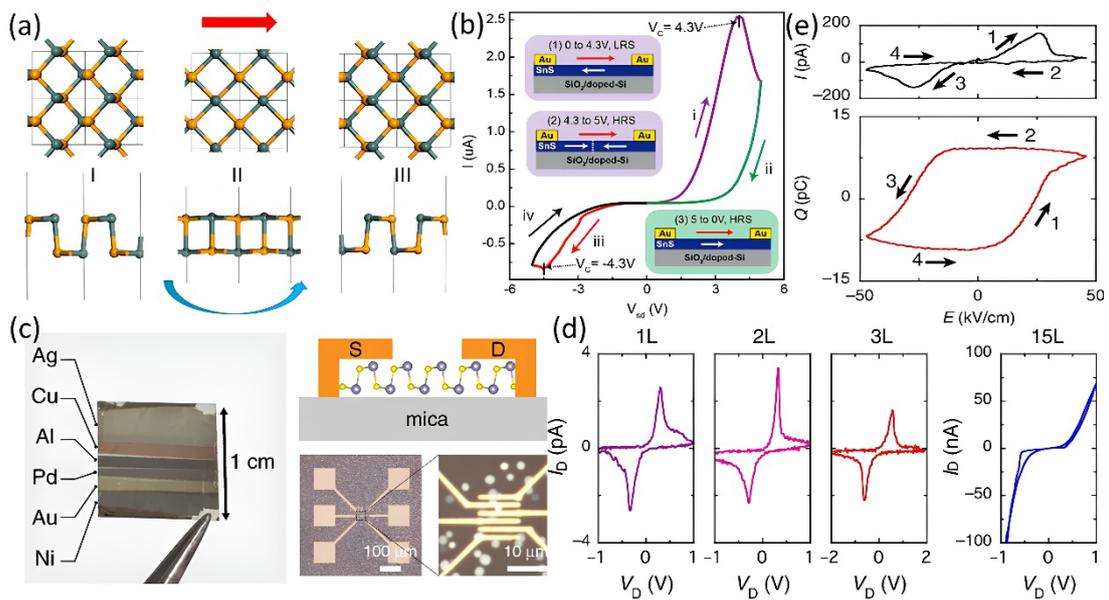

**Figure 16.** (a) Pathway of ferroelectric switching for a single-layer SnX' from top view (top panels) and side view (bottom panels). The red arrow denotes the direction of applied electric field while the blue arrow shows the transition from state I to state III. Reprinted with permission from [200]. Copyright 2016 American Chemical Society. (b) Current-voltage (I-V) hysteresis curve of a Fe-FET, where 15 nm-thick SnS film was deposited on a SiO$_2$/Si substrate.

The voltage swept in the order of 0→5→0→-5→0 V, labeled by i, ii, iii, and iv, with a gate voltage of -50 V. The insets are schematics illustrating the ferroelectric switching process during voltage sweeping. Reprinted with permission from [189]. Copyright 2019 American Chemical Society. (c-e) Room-temperature ferroelectric behaviors in a two-terminal device made of few-layer SnS/mica. Reproduced with permission from [191]. Copyright 2020, Springer Nature. (c) Left: photograph of a mica substrate with deposition of various metals with different work functions. Right: cross-sectional schematic and optical images of the two-terminal SnS-based device. (d) Ferroelectric resistive switching for SnS with different numbers of layers (L). (e) Ferroelectric resistive switching for a 9-layer SnS/Ag device: current (I) and charge(Q) vs. nominal electric field (E) measured by a room-temperature ferroelectric measurement system at 1 Hz.

Single-layer SnX' have been identified as promising ferroelectric materials due to their lack of structural inversion symmetry and spontaneous in-plane polarization.[175,182,186,189,191,200] Wu and Zeng[200] elucidated the polarization switching mechanism of single-layer SnX' in Figure 16a: when an electric field is applied parallel to the armchair direction, positive and negative ions move in opposite directions. The resulting structural transformation switches the direction of Sn-X' dipole moments and polarization by 180 °. The ferroelectric polarizations of SnS and SnSe are $2.47\times10^{-10}$ Cm$^{-1}$ and $1.87\times10^{-10}$ Cm$^{-1}$, respectively, which are comparable to other 2D ferroelectric materials like MoTe$_2$ and WTe$_2$.[201,202] The Curie temperature ($T_c$) is the transition point where a material loses its ferroelectric behavior. The $T_c$ of layered SnS and SnSe is 1200 K and 326 K, respectively.[175] Barraza-Lopez et al.[203] successfully raised the $T_c$ of single-layer SnSe by 250 K by applying uniaxial tensile strain.

Few-layer SnS has attracted interest as a high-efficiency ferroelectric material for its chemical stability and high $T_c$. Ferroelectric devices based on few-layer SnS have shown strong responses, as exemplified in Figure 16. An SnS-based Fe-FET (Figure 16b) showed a clear hysteretic I-V response under a bias from -5 to 5 V. The coercive voltage where the device transitions from a low-resistance to a high-resistance state was ±4.3 V, corresponding to a coercive field of 10.7 kVcm$^{-1}$. Higashitarumizu et al.[191] designed a two-terminal device based on few-layer SnS/mica, as shown in Figure 16c, and its ferroelectricity measured at room temperature is displayed in Figure 16d. Hysteretic I-V characteristics are observed in SnS films with a thickness of up to 15 layers. Interestingly, although ferroelectricity is theoretically expected to exist only in odd-layer systems, ferroelectric switching has been observed in bilayer-SnS devices because the stacking of SnS single layers can be tuned by external strain induced by the substrate. An optimized 9-layer SnS device (Figure 16e) presented a coercive electric field of 25 kVcm$^{-1}$.

## 6. Conclusion, Challenges & Outlook

In conclusion, research interest in nanolayered PTMMC semiconductors has grown exponentially in recent years, as their variety and tailored outstanding properties qualify them as attractive candidates for the next generation of nanoscale electronics and optics. We provide a brief summary and comparison of the applications of various PTMMC materials in several major areas as below:

2D GaX and InSe have been applied as channel materials in nanoscale FETs. Among them, GaTe exhibits excellent field-effect mobility and on/off ratio due to its relatively high charge carrier mobility and density, surpassing the extensively studied MoS$_2$-based FETs. GaTe and GeX' have the potential to be used in fabricating efficient photodetectors, as they show narrow bandgaps and low direct recombination rates. The resulting photodetectors have high photo-responsivity, fast response time, and moderate photo-detectivity. While GaX' and InSe have also been explored as photodetectors, their bandgap structures are not as suitable as GaTe and GeX'. In thermoelectric devices, SnX' are the most attention-worthy candidates, as the carrier mobilities of 2D SnX' are higher than most layered PTMMCs, though materials like γ-GeSe and GaX also show promise. In the field of nanoscale solar cells and other PV devices, GeX' (particularly γ-GeSe) nanosheets are favored for their extremely high carrier concentrations, low direct recombination rates, low resistance, strong light absorption, air stability, and high lattice flexibility. Furthermore, GaSe plays an important role in optical devices such as IR detectors and UV sensors, as it possesses nonlinear optical behaviors, high transparency in the IR spectrum, and γ'-GaSe exhibits unique resonant UV absorption. SnX', γ-GeSe, and InSe, which lack structural inversion symmetry, are good ferroelectric materials and have been applied in Fe-FETs. Also, GeX' are predicted to have high electrocatalytic potential, while GeSe outperforms GaS in electrochemical applications due to its better band edge position and higher photo absorption.

However, the applications of nanolayered PTMMCs mentioned above still have some shortcomings. Here, we outline their current challenges and possible solutions:

(1) GaX/InSe-based FETs still exhibit unsatisfactory carrier mobilities and on/off ratios. By exploring high-K dielectric materials as the top gate to fabricate top-gate FETs, it is possible to realize GaX/InSe-based FETs with much higher carrier mobility and on/off ratio. Developing polymers as dielectric materials and encapsulating channel materials for protection are also reliable ways to enhance performance.

(2) To further improve photodetector performance, one can design bottom-contact photodetectors to replace traditional top-contact configurations. Doping engineering is a common pathway to improve carrier mobility and density.

(3) Although SnX' and γ-GeSe are high-quality thermoelectric materials with large ZT coefficients, their high carrier concentrations hinder thermoelectric applications. Measures such as doping and alloying can effectively increase the Seebeck coefficient. On the other hand, SnS/SnSe-based thermoelectric devices show low room-temperature ZT values, which can be improved by nano-structuring.

(4) SnX', γ-GeSe, and InSe are ferroelectric semiconductors. Introducing strain can adjust their Curie temperatures and magnetic doping can overcome challenges posed by a large number of free charge carriers.

(5) Exfoliation is challenging for large-scale production of pure 2D single-crystal SnX'. MBE has been proven effective for synthesizing wafer-scale GaSe, InSe, and SnX' 2D crystalline films. However, it is not appropriate for GaS and GeX'. Other methods, such as CVD and PVD, are also under exploration.

(6) Devices based on materials like InSe and GaSe often face issues such as degradation and long-term instability in ambient conditions. Encapsulation can effectively solve these issues and extend the lifespan of devices.

**Author Contributions**

The manuscript was written through contributions of all authors. All authors have given approval to the final version of the manuscript.

ACKNOWLEDGMENT

This study is based upon research conducted at the Pennsylvania State University Two-Dimensional Crystal Consortium – Materials Innovation Platform which is supported by NSF cooperative agreement DMR-2039351. This research was partially supported by NSF through the University of Delaware Materials Research Science and Engineering Center, DMR-2011824. M. Y. and S. L. acknowledge funding from the Coherent/II-VI Foundation.

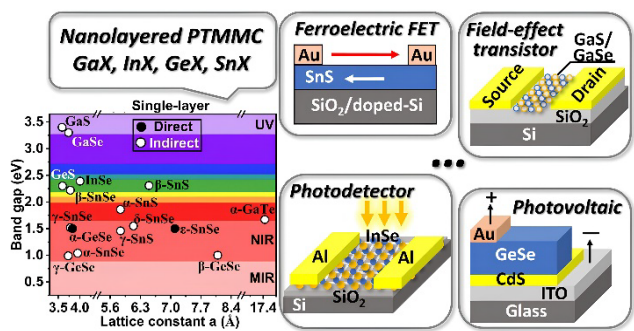

**ToC Figure**